\documentclass[12pt]{article}
\usepackage{amsmath,amssymb,epsfig}
\begin{document}
\parindent=0pt
\parskip=6pt
\rm

\vspace*{0.5cm}

\begin{center}
{\bf Dimensional crossover of the thermodynamic properties near the
phase transition to superconducting state in type I superconductors}

\vspace{0.3cm}

 D. V. Shopova$^{\ast}$, T. P. Todorov$^{\dag}$

{\em  CPCM Laboratory, Institute of Solid State Physics,\\
 Bulgarian Academy of Sciences, BG-1784 Sofia, Bulgaria.} \\
\end{center}

$^{\ast}$ Corresponding author.\\ {\em E-mail address:} sho@issp.bas.bg

$^{\dag}$ Permanent address: Joint Technical College at the Technical
University of Sofia.

\vspace{1cm}


\vspace{0.2cm}


\vspace{0.2cm}
\begin{abstract}
A new approach to the treatment of magnetic fluctuations in thin films
of type I superconductors is introduced. Results for the dependence of
 free energy,  specific heat and  order parameter profile on the
film thickness near the equilibrium phase transition point are obtained
and discussed.
\end{abstract}

Recent studies~\cite{Folk:2001,Shopova:2002,Shopova:2003} have shown
that the Halperin-Lubensky-Ma effect (HLM)~\cite{Halperin:1974},
namely, the fluctuation-driven change of the order of superconducting
phase transition in a zero magnetic field is much stronger in
quasi-two-dimensional (quasi-2D) type I superconducting films than in
bulk (3D) samples. This fact gives new opportunities for an
experimental verification of the effect in suitably chosen samples of
thin superconducting films. Numerical values of thermodynamic
quantities  as latent heat, specific heat capacity and order parameter
jump, have been calculated theoretically with the aim to support future
experiments~\cite{Shopova:2003}.

HLM effect occurs as a result of the interaction between the
superconducting order parameter $\psi$ and the vector potential
$\vec{A}(\vec{x})\equiv \delta \vec{A}(\vec{x})$ corresponding to the
fluctuation part of the magnetic induction $\vec{B} =\delta \vec{B}$;
note, that the mean value $B_0$ of the magnetic induction is assumed to
be zero. When this interaction is neglected the superconducting phase
transition in a zero external magnetic field is of second order. Taking
into account the magnetic fluctuations $\vec{A}(\vec{x})$ and their
gauge-invariant interaction with $\psi$ leads to an effective
Ginzburg-Landau (GL) free energy that describes a first order phase
transition at the phase transition point $T_{c0} = T_c(B_0 =0)$. In
type I superconductors the effect is stronger than in type II
superconductors and can be investigated theoretically with the help of
a self-consistent (mean-field-like) approximation, in which the scalar
field $\psi$ can be considered uniform, i.e., independent of the
spatial vector $\vec{x}$. Here we shall use this approximation; a more
detailed justification of this approach can be found in
Refs.~\cite{Folk:2001, Halperin:1974}. The coupling between a scalar
field, like $\psi$ in superconductors, and a vector gauge field, like
$\vec{A}$, is  present also in  models that describe other physical
systems, for example, in scalar electrodynamics, certain liquid
crystals and early universe phase transitions; see, e.g.,~\cite{
Folk:1999} for a recent review. The problem is relevant also to quantum
phase transitions in superconductors~\cite{Fisher:1988,Shopova2:2003}
and other systems~\cite{Belitz:2002}.

In this letter we investigate  the dimensional (2D-3D) crossover of
phase transition properties of  type I superconducting films. Our
treatment is based on preceding investigations of HLM effect
~\cite{Folk:2001,Shopova:2003}, where we have derived an effective
energy of the superconducting film from the Ginzburg-Landau model. We
consider the effective free energy~\cite{Shopova:2003} and related
thermodynamic quantities in the framework of a reliable and relatively
simple variant of the crossover theory, used in the study of disordered
films~\cite{Craco:1999}.
  Because of considerable differences, this method
can be applied to type I superconducting films only after an essential
modification~\cite{Rahola:2001}. Here we outline a new general way of
calculation of the thermodynamic functions dependence on the thickness
of  superconducting films. The present results, as summarized and
discussed at the end of this paper, confirm and extend the recent
investigations~\cite{Folk:2001,Shopova:2002,Shopova:2003} of the phase
transition properties in thin films of type I superconductors. Our
notations follow those in Ref.~\cite{Lifshitz:1980}.

We consider the following effective free energy density $f(\psi) =
F(\psi)/V$ of a thin superconducting film of thickness $L_0$ and volume
$V = (L_1 L_2L_0)$ ~\cite{Folk:2001, Halperin:1974}:
\begin{equation}
\label{eq1}
 f(\psi) = a|\psi|^2 + \frac{b}{2}|\psi|^4 +
k_BTJ\left[\rho\left(\psi\right)\right]\;,
\end{equation}
where
\begin{equation}
\label{eq2}
 J(\rho) = \int_0^{\Lambda}\frac{dk}{2\pi}k
  S\left(k,\rho \right);,
\end{equation}
\begin{equation}
\label{eq3}
 S = \frac{1}{L_0} \sum_{k_0 = -\Lambda_0}^{+
 \Lambda_0}\mbox{ln}\left[1 + \frac{\rho(\psi)}
 {k^2 +k^2_0}\right];.
\end{equation}
In Eqs.~(\ref{eq1})--(\ref{eq3}), $\rho(\psi) = \rho_0|\psi|^2$, where
$\rho_0 = (8\pi e^2/mc^2)$; $a = \alpha_0(T-T_{c0})$ and $b > 0$ are
the usual Landau parameters. They are related to the zero temperature
coherence length $\xi_0 = (\hbar^2/4m\alpha_0T_{c0})^{1/2}$, the
zero-temperature critical magnetic field $H_{c0} =
\alpha_0T_{c0}(4\pi/b)^{1/2}$, and the initial (unrenormalized)
critical temperature $T_{c0}$~\cite{Lifshitz:1980}. The third term in
the r.h.s. of Eq.~(\ref{eq1}) describes the effect of the magnetic
fluctuations~\cite{Folk:2001,Shopova:2003}. In
Eqs.~(\ref{eq2})--(\ref{eq3}), the integral $J(\rho)$ and the sum
$S(k,\rho)$ over the wave vector $\vec{q} = (\vec{k},k_0)$ are
truncated by the upper cutoffs $\Lambda$ and $\Lambda_0$. The finite
cutoff $\Lambda$ is introduced for the wave number $k=|\vec{k}|$ of the
wave vector $\vec{k}=(k_1,k_2)$ and $\Lambda_0$ stands for $k_0$. The
free energy ~(\ref{eq1})--(\ref{eq3})~ has been
derived~\cite{Folk:2001,Shopova:2003} under the condition $\kappa =
[\lambda(T)/\xi(T)] \ll 1/\sqrt{2}$, i.e. for type I superconductors.
Here, $\kappa$ is the Ginzburg-Landau number and the coherence length
$\xi$ is given by $\xi(T) = (\xi_0/\sqrt{t_0})$, where $t_0 =
[(T-T_{c0})/T_{c0}] \ll 1 $. The London penetration depth $\lambda(T)$
is  $\lambda(T) = (\lambda_0/\sqrt{t_0})$; $\lambda_0 =
(b/\rho_0\alpha_0T_{c0})^{1/2}$ corresponds to $T=0$.

In preceding investigations~\cite{Folk:2001,Shopova:2002,Shopova:2003}
of thin (quasi-2D) film, i.e., films obeying the condition $a_0 \ll L_0
\leq \Lambda^{-1}$,where $a_0$ is the lattice constant, only the
($k_0=0$)--term in the sum S has been taken into account. This
approximation leads to a simple dependence of the thermodynamic
quantities on the thickness $L_0$, but is not convenient when $L_0$
varies in broad limits. Here we shall apply a more general treatment
which gives the opportunity for the description of  2D-3D crossover
from quasi-2D to 3D-superconductors $L_0 \sim L_{\beta},\;
(\beta=1,2)$. The choice of the cutoff
 $\Lambda = (\pi/\xi_0)$ is consistent with the general
  requirement $\xi_0 < \lambda(T)$ for the validity of
	GL free energy for type I
superconductors (see, e.g., Ref.~\cite{Lifshitz:1980}) and  specific
requirements~\cite{Folk:2001,Shopova:2003} for the validity of the
Landau expansion of the free energy ~(\ref{eq1})--(\ref{eq3}) for
extremely thin films ($L_0 \ll \xi_0$).

As our study is based on the quasimacroscopic GL approach  the second
cutoff $\Lambda_0$ should be again related to $\xi_0$ rather than to
the lattice constant $a_0$, i.e. $\Lambda_0 \sim (1/\xi_0)$, which
means that phenomena at distances shorter than $\xi_0$ are excluded
from our consideration. We shall assume that the lowest possible value
of $\Lambda_0$ is $(\pi/\xi_0)$, as is for $\Lambda$, but we shall keep
in mind that both $\Lambda_0$ and $\Lambda$  can be extended to
infinity provided the main contribution to the integral $J(\rho)$ and
the sum $S$ come from the long wavelength limit $(q\xi_0\ll 1)$.

In a close vicinity of the phase transition point $T_{c0}$ from normal
($\psi = 0$) to Meissner state ($|\psi|
> 0$) the parameter $\rho \sim \psi^2$ is small and the main
contribution to the free energy $f$ will be given by the terms in $S$
with small wave vectors $ k \ll \Lambda$. This allows an approximate
but reliable treatment of the 2D-3D crossover by expanding the
summation over $k_0$ in~(\ref{eq3})  to infinity  - $\Lambda_0 \sim
\infty$; for a justification, see Ref.~\cite{Craco:1999, Rahola:2001}.
A variant of the theory when $\Lambda_0$ is kept finite ($\Lambda =
\Lambda_0 = \pi/\xi_0$) can also be developed but the results are too
complicated~\cite{Todorov:2003}. Performing the summation and the
integration in Eqs.~(\ref{eq2})--(\ref{eq3}) we obtain $J(\rho) =
(\Lambda^2/2\pi L_0)I(\rho)$, where
\begin{equation}
\label{eq4}
 I(\rho) = \int_0^{1}dy\:
 \mbox{ln}\left[\frac{\mbox
 {sh}\left(\frac{1}{2}\; L_0\Lambda\sqrt{\rho + y}\right)}{\mbox{sh}
 \left(\frac{1}{2}\; L_0\Lambda \sqrt{y}\right)}
  \right]\;,
\end{equation}
The integral~(\ref{eq4}) has a logarithmic divergence that corresponds
to the infinite contribution of  magnetic fluctuations to the free
energy of  normal phase ($ T_{c0} > 0, \varphi = 0$). Such  type of
divergence is a common property of a lot of phase transition models. In
the present case, as is in other systems, this divergence is
irrelevant, because the divergent term does not depend on the order
parameter $\psi$ and the free energy $f(\psi)$ is defined as the
difference between the total free energies of the superconducting and
normal phases: $f(\psi) = (f_S - f_N)$.

Introducing a dimensionless order parameter $\varphi = (\psi/\psi_0)$,
where $\psi_0 = (\alpha_0T_{c0}/b)^{1/2}$ is the value of $\psi$ at
$T=0$, we obtain the free energy~(\ref{eq1}) in the form
 \begin{equation}
\label{eq5}
 f(\varphi) = \frac{H_{c0}^2}{8\pi}\left[2t_0\varphi^2 + \frac{b}{2}|\varphi|^4 +
2(1+t_0)CI(\mu\varphi^2)\right]\;,
\end{equation}
with $I(\mu \varphi^2)$  given by Eq.~(\ref{eq4}), $\mu = (1/\pi
\kappa)^2$, $\Lambda = \pi/\xi_0$, and
\begin{equation}
\label{eq6}
 C = \frac{2\pi^2 k_BT_{c0}}{L_0\xi_0^2H^2_{c0}}\:.
\end{equation}
From the equation of state ($\partial f/\partial \varphi = 0$) we find
two possible phases: $\varphi_0 = 0$ and the superconducting phase
$(\varphi_0 > 0)$ defined by the equation
\begin{equation}
 \label{eq7}
 t_0 + \varphi^2 + \frac{(1+t_0)CL_0\xi_0}{4\pi \lambda_0^2}K(\mu\varphi^2) =
 0 \:,
\end{equation}
where
\begin{equation}
\label{eq8}
 K(z) = \int_0^{1}dy\:\frac{\mbox
 {coth}\left(\frac{1}{2}\;L_0\Lambda\sqrt{y + z}\right)}
 {\sqrt{y + z}}\:.
\end{equation}
The analysis of the stability condition ($\partial^2f/\partial\varphi^2
\ge 0$) shows that the normal phase is a minimum of $f(\varphi)$ for
$t_0 \geq 0$, whereas the superconducting phase is a minimum of
$f(\varphi)$ provided
\begin{equation}
\label{eq9}
 1 > \frac{1}{4}(1+t_0)CL_0\Lambda\mu^2\tilde{K}(\mu \varphi_0^2)\:,
 \end{equation}
 where
 \begin{equation}
 \label{eq10}
 \tilde{K}(z) = \int_0^{1}\frac{dy}{y + z}\left[\frac{\mbox
 {coth}\left(\frac{1}{2}\;L_0\Lambda\sqrt{y + z}\right)}
 {\sqrt{y + z}} + \frac{L_0\Lambda}{2\mbox{sh}^2\left(\frac{1}{2}\;L_0\Lambda\sqrt{y
 + z}\right)} \right]\:.
\end{equation}
  The entropy jump $\Delta s = (\Delta S/V) = [-df(\varphi_0)/dT]$ per unit volume at the
  equilibrium point of the phase transition $T_c \neq T_{c0}$ is obtained in the form
\begin{equation}
\label{eq11}
 \Delta s(T_c) = - \frac{H_{c0}^2\varphi_{c0}^2}{4\pi T_{c0}}
 \left[1+\frac{CI(\varphi_{c0})}
 {\varphi_{c0}^2}
 \right] \:,
\end{equation}
where $\varphi_{c0} \equiv \varphi_0(T_c)$ is the jump of the
dimensionless order parameter at $T_c$. The second term in $\Delta s$
can be neglected. In fact, taking into account the equation
$f[\varphi_0(T_c)] = 0$ for the equilibrium phase transition point
$T_c$  we obtain that $|CI(\varphi_0)/\varphi_0^2|$ is approximately
equal to $|t_{c0} + \varphi_{c0}^2/2|$, where $\varphi_{c0}^2$ and the
dimensionless shift of the transition temperature $t_{c0} = t_0(T_c)$
are expected to be much smaller than unity. The latent heat $Q =
T_c\Delta s(T_c)$ and the jump of the specific heat capacity at $T_c$,
$\Delta C = T_c(\partial \Delta S/\partial T)$ can be easily calculated
with the help of Eq.~(\ref{eq11}). For this purpose we need the
function $\varphi_0(T)$, which cannot be  obtained analytically from
Eq.~(\ref{eq7}).

Even the relatively simple 2D-3D crossover scheme outlined in this
paper cannot be used for an analytical calculation of the
superconducting order parameter $\varphi_(T)$ and related thermodynamic
quantities. But the simplicity of the scheme makes it very convenient
for a numerical calculation of experimentally observable quantities in
thin superconducting films. Our numerical treatment of
Eqs.~(\ref{eq5})~and~(\ref{eq7}) shows a remarkably good confirmation
of the results for 3D~\cite{Shopova:2002} and
quasi-2D~\cite{Folk:2001,Shopova:2003} superconductors. Here we want to
mention that  the reliability of present crossover scheme can be tested
also analytically by considering the cases: $L_0\Lambda \gg 1$, which
corresponds to the ``3D limit", and $L_0\Lambda \ll 1$, which gives the
``quasi-2D limit". After taking the quasi-2D limit in
Eqs.~(\ref{eq5})~and~(\ref{eq7}), we obtain the preceding results for
$\varphi_0(T)$ and the free energy $f(\varphi)$ reported in
Refs.~\cite{Folk:2001,Shopova:2003}. The results in the 3D limit of
Eqs.~(\ref{eq5})~and~(\ref{eq7}) should be compared with the results
for the free energy and the equation of state obtained when the
continuum limit ($\sum_k \rightarrow  \int_k $) in the sum~(\ref{eq3})
is taken, which means to extend  the integration  to $\Lambda_0 =
\infty$. In fact, we have done the calculations in the  geometry of  an
``infinite cylinder" that corresponds to our finite-size treatment.
Alternatively, the 2D-3D crossover can be described in the usual
``spherical" geometry, when the cutoffs $\Lambda$ and $\Lambda_0$ are
equal  both in  the sum~(\ref{eq3}) and in the respective integral,
calculated in the continuum limit ($L_0\Lambda \rightarrow \infty$)
along the small dimension $L_0$. The equations for the free energy and
the equation of state in this variant of the theory are much more
complicated and less convenient for a numerical evaluation of
experimentally observable thermodynamic quantities.

{\bf Acknowledgements.} The authors thank Prof. D. I. Uzunov for useful
discussions. A funding by the European Superconductivity Network
(Scenet) is acknowledged.




\end{document}